\documentclass[12pt]{article}
\usepackage{oldgerm}
\usepackage{euler}
\usepackage{amsmath}
\usepackage{amssymb}
\usepackage{amstext}
\usepackage{amscd}
\usepackage{amsfonts}
\usepackage{appendix}
\DeclareMathAlphabet{\EuFrak}{U}{euf}{m}{n}
\DeclareMathAlphabet{\EuScript}{U}{eus}{m}{n}

\newcommand{\be}{\begin{equation}}
\newcommand{\ee}{\end{equation}}
\newcommand{\ben}{\begin{eqnarray}}
\newcommand{\een}{\end{eqnarray}}

\title{{\bf New Solution of Diffusion-Advection
Equation for Cosmic-Ray Transport Using Ultradistributions}}
\author{M. C. Rocca$^{1,2}$,A. Plastino$^{1,2}$,\\ 
A. R. Plastino$^{3}$, A. L. DePaoli$^{1,2}$ \\
\small{$^1$Departamento de F\'{\i}sica, Fac. de Ciencias Exactas},\\
\small{Universidad Nacional de La Plata}\\
\small{C.C. 67 (1900) La Plata, Argentina}\\
\small{$^2$IFLP-CCT-CONICET-C.C 727 (1900) La Plata. Argentina}\\
\small{$^3$ CeBio y Secretaria de Investigacion,}\\
\small{Univ. Nac. del Noroeste de la Prov. de Bs. As.,}\\
\small{ UNNOBA and CONICET, R. Saenz Pe\~{n}a 456, Junin, Argentina}}

\date{\today}

\begin{document}

\maketitle

\begin{abstract}

In this paper we exactly solve the diffusion-advection
equation for cosmic-ray transport.
With this purpose we use the Theory of Ultradistributions
of J. Sebastiao e Silva, to give a general solution for this
equation. From this solution, we obtain several approximations as
limiting cases of various situations of physical and astrophysical
interest. One of them involves Solar cosmic-rays' diffusion.

Keywords: cosmic rays; diffusion; ultradistributions; fractional derivatives
\end{abstract}

\newpage

\renewcommand{\theequation}{\arabic{section}.\arabic{equation}}

\section{Introduction}

\subsection{The problem at hand}

Fractional derivatives constitute a  rather old subject, although
not as familiar as the integer-order counterparts \cite{uno,dos}.
Fractional derivatives have recently been used in regards to many
physical  problems [for a small sample, see for instance
\cite{barkai1,barkai2,tres,cuatro}] and  to hydrology \cite{cinco}.
Fractional derivatives have been recently applied to model
 super-diffusion of particles in astrophysical scenarios
\cite{LE2014,arxiv2014}. There is a
considerable evidence emerging  from data gathered  by spacecrafts
showing that the transport of energetic particles in the turbulent
heliospheric medium is super-diffusive \cite{PZ2007,PZ2009}. An
interesting work to be mentioned, in a different vein, is that of
\cite{seis}

People employ fractional space derivatives so as to model
anomalous diffusion or dispersion. Here, a particle plume spreads
at a rate that is not the one of a classical Brownian model. If a
fractional derivative takes the place of  the second derivative in
a diffusion or dispersion equation, this results in enhanced
diffusion, or
 super-diffusion. In the case of a constant coefficients,  one dimensional
advection-dispersion equation, analytical solutions can be found
by recourse to Fourier transform methods \cite{cinco}. Many other
problems require, instead,  a treatment with variable coefficients
\cite{ocho}.

In astrophysics, great  activity revolves around  the development
of super diffusive models for the transport of electrons and
protons in the heliosphere \cite{SS2011,TZ2011,ZP2013}. This sort
of transport  displays  a power-law growth of the mean square
displacement of the diffusing particles, $\langle \Delta x^2
\rangle \propto t^{\alpha}$, with $\alpha > 1$ (see, for instance,
\cite{SZ97}).
The special case $\alpha = 2$ is called ballistic. The limit case $\alpha \to 1$ is that of  normal
diffusion, described by a Gaussian propagator. Particles
associated with violent solar events  diffuse in the solar wind, a
turbulent scenario that can be taken as statistically homogeneous
at large  distances from the sun \cite{PZ2007}. This entails that
the propagator $P(x,x',t,t')$, describing the probability of
finding a particle that has been injected at $(x',t')$ at the
space time location $(x,t)$, depends solely on the differences
$x-x'$ and $t-t'$. In the super diffusive regime the propagator
$P(x,x',t,t')$ is not Gaussian, and is characterized by power-law
tails, emerging as the solution a non local diffusive process,
governed by an integral equation. This equation can be cast as a
diffusion one, in which the  Laplacian is replaced by a term
involving fractional derivatives \cite{C95}. See also
\cite{LTRLGS2014,
LSSEL2009,RLEL2007,stern,LMASL2005,MK2000,P2013}, and references
therein. An interesting  step towards a more accurate analytical
treatment of this problem was recently provided by Litvinenko and
Effenberger (LE) in \cite{LE2014}.

\subsection{Ultradistributions}

A series of papers \cite{tp1,tp2,tp3,tp4,tp5}  show that the
Ultradistribution theory of Sebastiao e Silva \cite{tp6,tp7,tp8}
permits a significant advance in the treatment of quantum field
theory. In particular, with the use of the convolution of
Ultradistributions, one can show that it  is possible to define a
general product of distributions (a product in a ring with
divisors of zero) that sheds new light on  the question of the
divergences in Quantum Field Theory. Furthermore,
Ultradistributions of Exponential Type (UET) are  adequate to
describe Gamow States and exponentially increasing fields in
Quantum Field Theory \cite{tp9,tp10,tp11}.

Other papers (\cite{tq1,ts2,ts3})  demonstrated that
Ultradistributions of Exponential type provide an adequate
framework for a consistent treatment of string and string field
theories. In particular, a general state of the closed string is
represented by UET of compact support, and as a consequence the
string field is a linear combination of UET of compact support.
Moreover,  five recent papers (\cite{tw1,tw2,tw3,tw4,tw5}) show
that Ultradistributions can be used to develop in a consistent way
the so-called Non-Extensive Statistical Mechanics, allowing for an
adequate definition of q-Fourier and q-Laplace transforms, and for
the removal of divergences of this theory.

Ultradistributions also have the advantage of being representable
by means of analytic functions. In general, they are easy to work
with  and, as we shall see, have interesting properties. One of
those properties is that Schwartz's tempered distributions are
canonical and continuously injected into Ultradistributions
Another interesting property is that the space of UET is reflexive
under the operation of Fourier transform (in a similar way of
tempered distributions of Schwartz)

\subsection{Our goal}

In this paper we wish to show that Ultradistributions provide an
adequate tool for a consistent treatment of a fractional
differential diffusion-advection equation.

A more conventional treatment of this equation is given in
\cite{arxiv2014}. The present treatment is of a much more general
character.

This paper is organized as follows: In section 2, we summarize a
set of mathematical concepts, while, in section 3 we formulate the
problem to be addressed herein. We obtain in Section 4 a general
solution of the fractional diffusion-advection equation, our main
result . In section 5, we discuss the so-called weak diffusion
approximation and in section 6 we analyze an important  change of
variables. Some conclusions are drawn in Section 7. The appendices
A, B, C and D give the fundamentals of the mathematical theory used
in this work.

\section{Some basic ideas related to Hilbert spaces}

The main task in distribution theory is to reinterpret functions
as linear functionals acting on a space of test functions. While
standard functions act by integration against a test function,
many other linear functionals do not arise in this way. Precisely,
 these are the {\it generalized functions}. One has a panoply of
possible choices for the space of test functions, which leads to
distinct spaces of distributions. The basic space of test
functions is that of smooth functions with compact support, which
yields standard distributions. Employing the space of smooth,
rapidly decreasing test functions gives instead the so-called
tempered distributions. They  are very important because they have
a well-defined  Fourier transform. While every tempered
distribution is a distribution in the normal sense, the converse
is not true. Generally, the larger the space of test functions,
the more restrictive the notion of distribution.

A nuclear space is a topological vector space with many of the
nice properties of finite-dimensional vector spaces. The topology
on them can be construed  by recourse to a family of semi-norms
whose unit balls-radii  decrease rapidly. Vector spaces whose
elements are {\it smooth} tend to be  nuclear spaces. An example
of  nuclear space is provided by the set of smooth functions on a
compact manifold. All finite-dimensional vector spaces are
nuclear.
 There are no Banach spaces that are nuclear, except for the
finite-dimensional ones. A large part of the theory of nuclear
spaces was developed by Alexander Grothendieck
(see \cite{tw6}).

A rigged  Hilbert space, or Guelfand's triplet, is a
construction designed to connect the distribution and
square-integrable aspects of functional analysis. Rigged spaces
were introduced to study spectral theory and bring together bound
states and continuous spectra. The concept of rigged Hilbert space
provides for such a desideratum an abstract functional-analytic
framework. More precisely,  an equipped Hilbert space consists of
a Hilbert space $\boldsymbol{H}$ plus a subspace $\boldsymbol{\Phi}$ 
which is endowed with a
finer topology. This  is one for which the natural inclusion

\be  \boldsymbol{\Phi} \subseteq \boldsymbol{H},  \ee is continuous.  
One can assume  that
$\boldsymbol{\Phi}$ is dense in $\boldsymbol{H}$ for the Hilbert norm. 
We need also the
inclusion of dual spaces $\boldsymbol{H^*}$ in  $\boldsymbol{\Phi^*}$. 
The latter, dual to
$\boldsymbol{\Phi}$ in its test function topology, is realized as a space of
distributions or generalized functions, and the linear functionals
on the subspace $\boldsymbol{\Phi}$ of type $\phi \rightarrow <v\rightarrow,
\phi>$ for $v$ in $\boldsymbol{H}$, are faithfully represented as
distributions. Now, via application of  the Riesz representation
theorem one  identifies $\boldsymbol{H^*}$ with $\boldsymbol{H}$. 
Therefore, the definition
of rigged Hilbert space can be given in terms of a double
inclusion
\be \boldsymbol{\Phi}\subseteq \boldsymbol{H} \subseteq  \boldsymbol{\Phi^*}.  \ee 
The most important
illustrations  are those for which $\boldsymbol{\Phi}$  is a nuclear space.
This captures an abstract expression of the idea that $\boldsymbol{\Phi}$
consists of test functions and $\boldsymbol{\Phi^*}$ of the corresponding
distributions. For more details, see Appendices A and B.

\section{Formulation of the Problem}

\setcounter{equation}{0}

The authors of \cite{LE2014} have proposed the equation:
\begin{equation}
\label{e2.1} \frac {{\partial f}(x,t)} {\partial t}=\kappa\frac
{{{\partial}^{\lambda} f}(x,t)} {\partial |x|^{\lambda}}+a\frac {{\partial
f}(x,t)} {\partial x}+\delta(x)\;\;\;t>0
\end{equation}
where $t>0$, for the distribution function $f(x,t)$. In their
specific case, $f$ refers to solar cosmic-ray transport. They used
the following definition of fractional derivative (see
\cite{SZ97}):
\begin{equation}
\label{e2.2} \frac {{\partial}^{\lambda} f} {\partial
|x|^{\lambda}}= \frac {1} {\pi}\sin\left(\frac {\pi\lambda}
{2}\right)\Gamma(\lambda+1) \int\limits_0^{\infty}\frac
{f(x+\xi)-2f(x)+f(x-\xi)} {{\xi}^{\lambda+1}}\;d\xi.
\end{equation}
To solve (\ref{e2.1}) the authors use the Green function given by
\begin{equation}
\label{e2.3} \frac {{\partial {\cal G}}(x,t)} {\partial t}=\kappa\frac
{{{\partial}^{\lambda} {\cal G}}(x,t)} {\partial
|x|^{\lambda}}+\delta(x)\delta(t).
\end{equation}
Using this Green function, the solution of (\ref{e2.1}), with the
initial condition $f(x,0)=0$, can be written as
\begin{equation}
\label{e2.4} f(x,t)=\int\limits_0^t {\cal G}(x+at^{'},
t^{'})\;dt^{'}.
\end{equation}
The solution to the above problem is well posed, except for one
major problem: the fractional derivative used is not defined for
$\lambda=1$ and {\it does not coincide for this value of $\lambda$
with the usual derivative defined by Newton and Leibniz.}

We will solve in this paper this serious problem by recourse to a
definition of fractional derivative {\it valid for all values of
$\lambda$, both real or complex}, and matching things for
$\lambda\in{\cal N}$ (${\cal N}$= the set of natural numbers),
with the usual derivative defined by Newton and Leibniz. To
achieve this goal we use the definition given in \cite{tp1} for
distributions of exponential type and extended in our Appendix C
to ultradistributions of exponential type.

An interesting property of this fractional derivative is that it
unifies in a single operation the  operations of derivation and
indefinite integration, for any real or complex value of
$\lambda$.

\section{General Solutions}

\setcounter{equation}{0}

To solve (\ref{e2.1}) we divide the problem into two parts:\\
$1) x\geq 0$
\begin{equation}
\label{e3.1} \frac {{\partial f}(x,t)} {\partial t}=\kappa\frac
{{{\partial}^{\lambda} f}(x,t)} {\partial x^{\lambda}}+a\frac {{\partial
f}(x,t)} {\partial x}+\delta(x),
\end{equation}
\begin{equation}
\label{e3.2} \frac {{\partial {\cal G}}(x,t)} {\partial t}=\kappa\frac
{{{\partial}^{\lambda} {\cal G}}(x,t)} {\partial
x^{\lambda}}+\delta(x)\delta(t).
\end{equation}
and\\
$2)x<0$
\begin{equation}
\label{e3.3} \frac {{\partial f}(x,t)} {\partial t}=\kappa\frac
{{{\partial}^{\lambda} f}(x,t)} {\partial (-x)^{\lambda}}+a\frac {{\partial
f}(x,t)} {\partial x}+\delta(x),
\end{equation}
\begin{equation}
\label{e3.4} \frac {{\partial {\cal G}}(x,t)} {\partial t}=\kappa\frac
{{{\partial}^{\lambda} {\cal G}}(x,t)} {\partial
(-x)^{\lambda}}+\delta(x)\delta(t).
\end{equation}
Our solution will be valid for all values of $\lambda$ such that
$|e^{\kappa(-ik)^{\lambda}}|\leq |e^{\kappa k}|$. For the
remaining possible values of $\lambda$, the solution is obtained
via analytic prolongation. In fact, for these last values, the
solutions of the  equations above become exponentially growing
ones, which forces one to i) appeal  to  ultradistributions of
exponential type and ii)  extend these  equations to the complex
plane. Thus, we have for $1)$
\begin{equation}
\label{e3.5} \frac {{\partial f}(z,t)} {\partial t}=\kappa\frac
{{{\partial}^{\lambda} f}(z,t)} {\partial z^{\lambda}}+a\frac {{\partial
f}(z,t)} {\partial z}+\delta(z),
\end{equation}
\begin{equation}
\label{e3.6} \frac {{\partial {\cal G}}(z,t)} {\partial
t}=\kappa\frac {{{\partial}^{\lambda} {\cal G}}(z,t)} {\partial
z^{\lambda}}+\delta(z)\delta(t),
\end{equation}
and for $2)$
\begin{equation}
\label{e3.7} \frac {{\partial f}(z,t)} {\partial t}=\kappa\frac
{{{\partial}^{\lambda} f}(z,t)} {\partial (-z)^{\lambda}}+a\frac {{\partial
f}(z,t)} {\partial z}+\delta(z),
\end{equation}
\begin{equation}
\label{e3.8} \frac {{\partial {\cal G}}(z,t)} {\partial t}=\kappa\frac
{{{\partial}^{\lambda} {\cal G}}(z,t)} {\partial
(-z)^{\lambda}}+\delta(z)\delta(t).
\end{equation}
Using now the complex Fourier transformation we can obtain the
solution to our four equations. For $1)$ one has
\[f(z,t)=\frac {1} {2\pi}\oint\limits_{\Gamma}
\{H[\Im(z)]H[-\Re(k)]-H[-\Im(z)]H[\Re(k)]\}H[\Im(k)]\times \]
\begin{equation}
\label{e3.9}
\frac {e^{[\kappa(-ik)^{\lambda}-iak]t}-1}
{\kappa(-ik)^{\lambda}-iak}e^{-ikz}\;dk
\end{equation}
\[{\cal G}(z,t)=\frac {H(t)} {2\pi}\oint\limits_{\Gamma}
\{H[\Im(z)]H[-\Re(k)]-H[-\Im(z)]H[\Re(k)]\}H[\Im(k)]\times \]
\begin{equation}
\label{e3.10} e^{\kappa(-ik)^{\lambda}t}e^{-ikz}\;dk,
\end{equation}
and for $2)$
\[f(z,t)=-\frac {1} {2\pi}\oint\limits_{\Gamma}
\{H[\Im(z)]H[-\Re(k)]-H[-\Im(z)]H[\Re(k)]\}H[-\Im(k)]\times \]
\begin{equation}
\label{e3.11}
\frac {e^{[\kappa(ik)^{\lambda}-iak]t}-1}
{\kappa(-ik)^{\lambda}-iak}e^{-ikz}\;dk
\end{equation}
\[{\cal G}(z,t)=-\frac {H(t)} {2\pi}\oint\limits_{\Gamma}
\{H[\Im(z)]H[-\Re(k)]-H[-\Im(z)]H[\Re(k)]\}H[-\Im(k)]\times \]
\begin{equation}
\label{e3.12} e^{\kappa(ik)^{\lambda}t}e^{-ikz}\;dk.
\end{equation}
We pass now to find explicit expressions for equations
(\ref{e3.9})-(\ref{e3.12}).

\subsection*{Case $1)\; x\geq 0$}

Expanding $e^{\kappa(-ik)^{\lambda}t}$ in power series we obtain,
for (\ref{e3.10}),
\[{\cal G}(z,t)=\frac {H(t)} {2\pi}
\sum\limits_{n=0}^{\infty}\frac {{\kappa}^nt^n} {n!}
\oint\limits_{\Gamma}
\{H[\Im(z)]H[-\Re(k)]-H[-\Im(z)]H[\Re(k)]\}\times \]
\begin{equation}
\label{e3.13} H[\Im(k)](-ik)^{\lambda n}e^{-ikz}\;dk.
\end{equation}
Each term of the sum in (\ref{e3.13}) is a tempered
ultradistribution (see Appendix A). We go then  to the real axis
and evaluate the cut along it. Thus,
\begin{equation}
\label{e3.14} {\cal G}(x,t)=\frac {H(t)} {2\pi}
\sum\limits_{n=0}^{\infty}\frac {{\kappa}^nt^n} {n!}
\oint\limits_{\Gamma}H[\Im(k)](-ik)^{\lambda n}e^{-ikx}\;dk.
\end{equation}
Eq. (\ref{e3.14}) can be cast in the following equivalent form
\begin{equation}
\label{e3.15} {\cal G}(x,t)=\frac {H(t)} {2\pi}
\sum\limits_{n=0}^{\infty}\frac {{\kappa}^nt^n} {n!} e^{-i\frac
{\pi} {2}\lambda n} \int\limits_{-\infty}^{\infty}(k+i0)^{\lambda
n}e^{-ikx}\;dk,
\end{equation}
or
\begin{equation}
\label{e3.16} {\cal G}(x,t)=\frac {H(t)} {2\pi}
\sum\limits_{n=0}^{\infty}\frac {{\kappa}^nt^n} {n!} e^{-i\frac
{\pi} {2}\lambda n} \int\limits_0^{\infty}\left(k^{\lambda
n}e^{-ikx}+ e^{i\pi\lambda n}k^{\lambda n}e^{ikx}\right)\;dk.
\end{equation}
The integrals given in (\ref{e3.16}) have been calculated in
\cite{guelfand1}. One has
\begin{equation}
\label{e3.17} {\cal G}(x,t)=\frac {iH(t)} {2\pi}
\sum\limits_{n=0}^{\infty}\frac {{\kappa}^nt^n} {n!}
\Gamma(\lambda n+1)\left[ \frac {e^{i\pi\lambda n}}
{(x+i0)^{\lambda n+1}}- \frac {e^{-i\pi\lambda n}}
{(x-i0)^{\lambda n+1}}\right].
\end{equation}
By recourse to (\ref{e2.4}) we have for $f$
\[f(x,t)=\frac {i} {2\pi}
\sum\limits_{n=0}^{\infty}\frac {{\kappa}^n} {n!}
\Gamma(\lambda n+1)\int\limits_0^t\left[
\frac {e^{i\pi\lambda n}t^{'n}} {(x+at^{'}+i0)^{\lambda n+1}}-\right.\]
\begin{equation}
\label{e3.18} \left.\frac {e^{-i\pi\lambda n}t^{'n}}
{(x+at^{'}-i0)^{\lambda n+1}} \right]\;dt^{'}.
\end{equation}
Using (\ref{d1},) we obtain the general solution for $x\geq 0$
\[f(x,t)=\frac {i} {2\pi}
\sum\limits_{n=0}^{\infty}\frac {{\kappa}^nt^n} {n!}
\Gamma(\lambda n+1)B(1,n+1)\times \]
\[\left[
\frac {e^{i\pi\lambda n}} {(x+i0)^{\lambda n+1}}
F\left(\lambda n+1,n+1;n+2;-\frac {at} {x+i0}\right)-\right.\]
\begin{equation}
\label{e3.19} \left.\frac {e^{-i\pi\lambda n}} {(x-i0)^{\lambda
n+1}} F\left(\lambda n+1,n+1;n+2;-\frac {at} {x-i0}\right)
\right].
\end{equation}

\subsection*{Case $2)\; x<0$}

From (\ref{e3.12}),  and expanding $e^{(ik)^{\lambda}}$ in power
series, ${\cal G}$ adopts the form
\[{\cal G}(z,t)=-\frac {H(t)} {2\pi}
\sum\limits_{n=0}^{\infty}\frac {{\kappa}^nt^n} {n!}
\oint\limits_{\Gamma}
\{H[\Im(z)]H[-\Re(k)]-H[-\Im(z)]H[\Re(k)]\}\times \]
\begin{equation}
\label{e3.20} H[-\Im(k)](ik)^{\lambda n}e^{-ikz}\;dk.
\end{equation}
Each term of the sum in (\ref{e3.20}) is, again,  a tempered
ultradistribution. Thus, proceeding as in the case $x\geq 0$, we
obtain
\begin{equation}
\label{e3.21} {\cal G}(x,t)=-\frac {H(t)} {2\pi}
\sum\limits_{n=0}^{\infty}\frac {{\kappa}^nt^n} {n!}
\oint\limits_{\Gamma}H[-\Im(k)](ik)^{\lambda n}e^{-ikx}\;dk,
\end{equation}
or
\begin{equation}
\label{e3.22} {\cal G}(x,t)=\frac {H(t)} {2\pi}
\sum\limits_{n=0}^{\infty}\frac {{\kappa}^nt^n} {n!} e^{i\frac
{\pi} {2}\lambda n} \int\limits_{-\infty}^{\infty}(k-i0)^{\lambda
n}e^{-ikx}\;dk.
\end{equation}
Eq. (\ref{e3.22}) can be rewritten as
\begin{equation}
\label{e3.23} {\cal G}(x,t)=\frac {H(t)} {2\pi}
\sum\limits_{n=0}^{\infty}\frac {{\kappa}^nt^n} {n!} e^{i\frac
{\pi} {2}\lambda n} \int\limits_0^{\infty}\left(k^{\lambda
n}e^{-ikx}+ e^{-i\pi\lambda n}k^{\lambda n}e^{ikx}\right)\;dk.
\end{equation}
Using \cite{guelfand1} we then  have
\begin{equation}
\label{e3.24} {\cal G}(x,t)=\frac {iH(t)} {2\pi}
\sum\limits_{n=0}^{\infty}\frac {{\kappa}^nt^n} {n!}
\Gamma(\lambda n+1)\left[ \frac {1} {(x+i0)^{\lambda n+1}}- \frac
{1} {(x-i0)^{\lambda n+1}}\right].
\end{equation}
As it should,  ${\cal G}$ vanishes for $x\geq 0$. From
(\ref{e2.4}), we can write $f$ as
\[f(x,t)=\frac {i} {2\pi}
\sum\limits_{n=0}^{\infty}\frac {{\kappa}^n} {n!}
\Gamma(\lambda n+1)\int\limits_0^t\left[
\frac {t^{'n}} {(x+at^{'}+i0)^{\lambda n+1}}-\right.\]
\begin{equation}
\label{e3.25} \left.\frac {t^{'n}} {(x+at^{'}-i0)^{\lambda n+1}}
\right]\;dt^{'},
\end{equation}
and, according to (\ref{d1}), $f$ is finally given by
\[f(x,t)=\frac {i} {2\pi}
\sum\limits_{n=0}^{\infty}\frac {{\kappa}^nt^n} {n!}
\Gamma(\lambda n+1)B(1,n+1)\times \]
\[\left[
\frac {1} {(x+i0)^{\lambda n+1}}
F\left(\lambda n+1,n+1;n+2;-\frac {at} {x+i0}\right)-\right.\]
\begin{equation}
\label{e3.26} \left.\frac {1} {(x-i0)^{\lambda n+1}}
F\left(\lambda n+1,n+1;n+2;-\frac {at} {x-i0}\right) \right].
\end{equation}

\section{A useful approximation}

\setcounter{equation}{0}

Following \cite{LE2014}, we shall now consider a weak diffusion
approximation. Within this approximation, we can treat $\kappa$ as
a small parameter and expand $f$ up to order one \cite{LE2014}.
Thus, we can write for $x\geq 0$
\begin{equation}
\label{e4.1}
f(x,t)=f_0(x,t)+f_1(x,t),
\end{equation}
where $f_0$ is given by
\[f_0(x,t)=\frac {it} {2\pi}\left[
(x+i0)^{-1}F\left(1,1;2;-\frac {at} {x+i0}\right)\right.-\]
\begin{equation}
\label{e4.2}
\left.(x-i0)^{-1}F\left(1,1;2;-\frac {at}
{x-i0}\right)\right].
\end{equation}
By recourse to \cite{gra1}, we can express $F(1,1;2;z)$ in terms
of elementary functions, i.e.,
\begin{equation}
\label{e4.3}
F(1,1;2;-z)=\frac {1} {z}\ln(1+z),
\end{equation}
and obtain for $f_0$ the expression
\begin{equation}
\label{e4.4}
f_0(x,t)=\frac {1} {a}[H(-x)-H(-x-at)=\frac {1} {2a}
[Sgn(x+at)-Sgn(x)].
\end{equation}
For $f_1$ we have
\[f_1(x,t)=\frac {i\kappa t^2} {4\pi}\Gamma(\lambda+1)\left[
\frac {e^{i\pi\lambda}} {(x+i0)^{\lambda+1}}
F\left(\lambda+1,2;3;-\frac {at} {x+i0}\right)\right.-\]
\begin{equation}
\label{e4.5}
\left.\frac {e^{-i\pi\lambda}}
{(x-i0)^{\lambda+1}} F\left(\lambda+1,2;3;-\frac {at}
{x-i0}\right)\right].
\end{equation}
Using again the result (\ref{d7}) one has
\[F(\lambda+1,2;3;z)=\frac {2} {\lambda(\lambda-1)z^2}
\left[1+\frac {\lambda z-1} {(1-z)^{\lambda}}\right],\] $f_1$ and
then
\[f_1(x,t)=\frac {i\kappa\Gamma(\lambda-1)} {2\pi a^2}
\left\{e^{i\pi\lambda}
\left[\frac {1} {(x+i0)^{\lambda-1}}-
\frac {x+\lambda at} {(x+at+i0)^{\lambda}}\right]-
\right.\]
\begin{equation}
\label{e4.6}
\left. e^{-i\pi\lambda}
\left[\frac {1}
{(x-i0)^{\lambda-1}}- \frac {x+\lambda at}
{(x+at-i0)^{\lambda-1}}\right]\right\}.
\end{equation}
Thus, we have for $f$ in the weak diffusion approximation
\[f(x,t)=\frac {1} {2a}[Sgn(x+at)-Sgn(x)]+\]
\[\frac {i\kappa\Gamma(\lambda-1)} {2\pi a^2}
\left\{e^{i\pi\lambda}
\left[\frac {1} {(x+i0)^{\lambda-1}}-
\frac {x+\lambda at} {(x+at+i0)^{\lambda}}\right]-
\right.\]
\begin{equation}
\label{e4.7}
\left. e^{-i\pi\lambda}
\left[\frac {1}
{(x-i0)^{\lambda-1}}- \frac {x+\lambda at}
{(x+at-i0)^{\lambda}}\right]\right\}.
\end{equation}
For $x>0$ $f$ this becomes simplified and one has
\begin{equation}
\label{e4.8}
f(x,t)=\frac {\kappa}
{a^2\Gamma(2-\lambda)} \left[\frac {1} {x^{\lambda-1}}-
\frac {x+\lambda at} {(x+at)^{\lambda}}\right].
\end{equation}
We can now distinguish two limiting cases. The first one is the
asymptotic situation $x>>at$. In this instance
\begin{equation}
\label{e4.9}
f(x,t)=\frac {1} {2\Gamma(-\lambda)}
\frac {\kappa t^2} {x^{\lambda+1}}.
\end{equation}
The second case is $0<x<<at$. For it we have
\begin{equation}
\label{e4.10}
f(x,t)=\frac {1} {\Gamma(2-\lambda)}
\frac {\kappa} {a^2}x^{1-\lambda}.
\end{equation}
For $x<0$ we have for $f_0$  the same expression obtained  for the
case $x\geq 0$, and thus, for $f_1$,
\[f_1(x,t)=\frac {i\kappa t^2} {4\pi}\Gamma(\lambda+1)\left[
\frac {1} {(x+i0)^{\lambda+1}}
F\left(\lambda+1,2;3;-\frac {at} {x+i0}\right)\right.-\]
\begin{equation}
\label{e4.11}
\left.\frac {1}
{(x-i0)^{\lambda+1}} F\left(\lambda+1,2;3;-\frac {at}
{x-i0}\right)\right].
\end{equation}
As a consequence, we have for $f$
\[f(x,t)=\frac {1} {2a}[Sgn(x+at)-Sgn(x)]+\]
\[\frac {i\kappa\Gamma(\lambda-1)} {2\pi a^2}
\left[\frac {1} {(x+i0)^{\lambda-1}}-
\frac {1} {(x-i0)^{\lambda-1}}
\right.\]
\begin{equation}
\label{e4.12} \left. -\frac {x+\lambda at} {(x+at+i0)^{\lambda}}+
\frac {x+\lambda at} {(x+at-i0)^{\lambda}}\right].
\end{equation}
For  $x+at<0$, (\ref{e4.2}) adopts the form
\begin{equation}
\label{e4.13} f(x,t)=\frac {\kappa} {a^2\Gamma(2-\lambda)}
\left[\frac {1} {|x|^{\lambda-1}}+ \frac {x+\lambda at}
{|x+at|^{\lambda}}\right].
\end{equation}
When $x<<-at$, (\ref{e4.13}) transforms into
\begin{equation}
\label{e4.14}
f(x,t)=\frac {1} {2\Gamma(-\lambda)}
\frac {\kappa t^2} {|x|^{\lambda+1}}.
\end{equation}
Another special situation arises when $x<0$, $x+at>0$, and
$x<<-at$. In this case, from (\ref{e4.12}) we  deduce the
following expression for $f$:
\begin{equation}
\label{e4.15}
f(x,t)=\frac {1} {a}+
\frac {1} {\Gamma(2-\lambda)}
\frac {\kappa} {a^2}|x|^{1-\lambda}.
\end{equation}

\section{Change of frame}

\setcounter{equation}{0}

Assume that, in the solar wind rest frame, the particles'
transport is represented by the fractional-diffusion equation
(FDE) without advection term ($a=0$ in (\ref{e2.1})). The shock
front, started at $x_0 = -V_{sh} t_0$, moves with constant speed
$V_{sh}$. It is considered  as highly localized in the
$x$-coordinate) and constitutes the source of the particles. Then
we face an FDE with a uniformly moving Dirac's delta source of the
form $\delta(x - V_{sh} t)$. So as to have a stationary delta
source,  we require  performing  a suitable  coordinates-change,
reformulating our task in a reference frame where the shock front
is stationary. We also modify the time-origin  so that the source
begins being active at $t=0$. In such a modified  reference frame,
the transport equation acquires an advection term with  velocity
$a= V_{sh}$, and a stationary source $\delta (0)$, that begins at
$t=0$. After solving the diffusion-advection equation in this
frame, one  expresses the solution in terms of the original
coordinates associated with the solar wind rest frame. Such  step
is briefly  described by the 3 correspondences $a\rightarrow
v_{sh}$, $t\rightarrow t+t_0$, and $x\rightarrow x-v_{sh}t$.
Consequently, Eqs. (\ref{e3.19}) and (\ref{e3.26}) acquire the
form, for $x\geq 0$,
\[f(x,t)=\frac {i} {2\pi}\sum\limits_{n=0}^{\infty}
 \frac {{\kappa}^n(t+t_0)^{n+1}} {n!}\Gamma(\lambda n+1)
{\cal B}(1,n+1)\times\]
\[\left[\frac {e^{i\pi\lambda n}}
{(x-v_{sh}t+i0)^{\lambda n + 1}}
F\left(\lambda n+1,n+1;n+2;-\frac {v_{sh}(t+t_0)} {x-v_{sh}t+i0}\right)-\right.\]
\begin{equation}
\label{e5.1}
\left.\frac {e^{-i\pi\lambda n}}
{(x-v_{sh}t-i0)^{\lambda n + 1}} F\left(\alpha n+1,n+1;n+2;-\frac
{v_{sh}(t+t_0)} {x-v_{sh}t-i0}\right)\right].
\end{equation}
And for $x<0$:
\[f(x,t)=\frac {i} {2\pi}\sum\limits_{n=0}^{\infty}
 \frac {{\kappa}^n(t+t_0)^{n+1}} {n!}\Gamma(\lambda n+1)
{\cal B}(1,n+1)\times\]
\[\left[\frac {1}
{(x-v_{sh}t+i0)^{\lambda n + 1}}
F\left(\lambda n+1,n+1;n+2;-\frac {v_{sh}(t+t_0)} {x-v_{sh}t+i0}\right)-\right.\]
\begin{equation}
\label{e5.2}
\left.\frac {1}
{(x-v_{sh}t-i0)^{\lambda n + 1}} F\left(\alpha n+1,n+1;n+2;-\frac
{v_{sh}(t+t_0)} {x-v_{sh}t-i0}\right)\right].
\end{equation}
Thus,  in  the weak diffusion approach of (\ref{e4.7}) and
(\ref{e4.12}), we have for $x\geq 0$
\[f(x,t)=\frac {1} {2v_{sh}}
[Sgn(x+v_{sh}t_0)-Sgn(x-v_{sh}t)]-\]
\[\frac {i\kappa\Gamma(\lambda-1)} {2\pi V_{sh}^2}
\left\{(x+(\lambda-1) v_{sh}t+v_{sh}t_0)
\left[\frac {e^{i\pi\lambda}} {(x+v_{sh}t_0+i0)^{\lambda}}-
\right.\right.-\]
\begin{equation}
\label{e5.3} \left.\left.\frac {e^{-i\pi\lambda}}
{(x+v_{sh}t_0-i0)^{\lambda}}\right] +\frac {e^{-i\pi\lambda}}
{(x-v_{sh}t-i0)^{\lambda-1}}- \frac {e^{i\pi\lambda}}
{(x-v_{sh}t+i0)^{\lambda-1}}\right\}.
\end{equation}
For $x<0$ we have
\[f(x,t)=\frac {1} {2v_{sh}}
[Sgn(x+v_{sh}t_0)-Sgn(x-v_{sh}t)]-\]
\[\frac {i\kappa\Gamma(\lambda-1)} {2\pi V_{sh}^2}
\left\{(x+(\lambda-1) v_{sh}t+v_{sh}t_0)
\left[\frac {1} {(x+v_{sh}t_0+i0)^{\lambda}}-
\right.\right.-\]
\begin{equation}
\label{e5.4} \left.\left.\frac {1}
{(x+v_{sh}t_0-i0)^{\lambda}}\right] +\frac {1}
{(x-v_{sh}t-i0)^{\lambda-1}}- \frac {1}
{(x-v_{sh}t+i0)^{\lambda-1}}\right\}.
\end{equation}

\section{Conclusions}

\setcounter{equation}{0}

By recourse to ultradistributions, we have provided  here an
explicit analytical solution for an advection-diffusion equation
(ADE) involving fractional derivatives. First, we devised a
generalized treatment for these derivatives that includes the
normal case.

 We also found the exact solution for the ADE both  in the $x$-configuration space and in  the
 associated $k$-space, that are related
  via a Fourier transform. Our solution
 allows us to obtain in a unified and systematic fashion all the
 different  approximations that were introduced
 in \cite{LE2014}, each one in a distinct manner. We achieve in
 this way a great degree of generality.

\newpage

\renewcommand{\thesection}{\Alph{section}}

\renewcommand{\theequation}{\Alph{section}.\arabic{equation}}

\setcounter{section}{1}

\section*{Appendix A}

\setcounter{equation}{0}

\subsection*{Distributions of Exponential Type}

For the benefit of the reader, we present here a brief description
of the main properties of Tempered Ultradistributions and of 
Ultradistributions of Exponential Type.

{\bf Notations}.
The notations are almost textually taken from ref\cite{tp7}.
Let $\boldsymbol{{\mathbb{R}}^n}$
(respectively $\boldsymbol{{\mathbb{C}}^n}$)
be the real (respectively complex)
n-dimensional space whose points are denoted by $x=(x_1,x_2,...,x_n)$
(resp $z=(z_1,z_2,...,z_n)$). We shall use the following notations:

(i) $x+y=(x_1+y_1,x_2+y_2,...,x_n+y_n)$\; ; \;
    $\alpha x=(\alpha x_1,\alpha x_2,...,\alpha x_n)$

(ii)$x\geqq 0$ means $x_1\geqq 0, x_2\geqq 0,...,x_n\geqq 0$

(iii)$x\cdot y=\sum\limits_{j=1}^n x_j y_j$

(iV)$\mid x\mid =\sum\limits_{j=1}^n \mid x_j\mid$

Consider the set of n-tuples of natural numbers $\boldsymbol{{\mathbb{N}}^n}$.
If $p\in\boldsymbol{{\mathbb{N}}^n}$, then
$p=(p_1, p_2,...,p_n)$,
where $p_j$ is a natural number, $1\leqq j\leqq n$. $p+q$ denote
$(p_1+q_1, p_2+q_2,..., p_n+q_n)$ and $p\geqq q$ means $p_1\geqq q_1,
p_2\geqq q_2,...,p_n\geqq q_n$. $x^p$ means $x_1^{p_1}x_2^{p_2}...
x_n^{p_n}$. We denote by
$\mid p\mid=\sum\limits_{j=1}^n  p_j $ and by $D^p$ we understand the
differential operator ${\partial}^{p_1+p_2+...+p_n}/\partial{x_1}^{p_1}
\partial{x_2}^{p_2}...\partial{x_n}^{p_n}$

For any natural number $k$ we define $x^k=x_1^k x_2^k...x_n^k$
and ${\partial}^k/\partial x^k=
{\partial}^{nk}/\partial x_1^k\partial x_2^k...\partial x_n^k$

The space $\boldsymbol{{\cal H}}$  of test functions
such that $e^{p|x|}|D^q\phi(x)|$ is bounded for any natural numbers
p and q is defined ( ref.\cite{tp7} ) by means
of the countably set of norms:
\begin{equation}
\label{a1}
{\|\hat{\phi}\|}_p=\sup_{0\leq q\leq p,\,x}
e^{p|x|} \left|D^q \hat{\phi} (x)\right|\;\;\;,\;\;\;p=0,1,2,...
\end{equation}
According to reference\cite{gue} $\boldsymbol{{\cal H}}$  is a
$\boldsymbol{{\cal K}\{M_p\}}$ space
with:
\begin{equation}
\label{a2}
M_p(x)=e^{(p-1)|x|}\;\;\;,\;\;\; p=1,2,...
\end{equation}
$\boldsymbol{{\cal K}\{e^{(p-1)|x|}\}}$ complies condition
$\boldsymbol({\cal N})$
of Guelfand ( ref.\cite{tp12} ). It is a countable Hilbert and nuclear
space:
\begin{equation}
\label{a3}
\boldsymbol{{\cal K}\{e^{(p-1)|x|}\}} =\boldsymbol{{\cal H}} =
\bigcap\limits_{p=1}^{\infty}\boldsymbol{{\cal H}_p}
\end{equation}
where $\boldsymbol{{\cal H}_p}$ is obtained by completing
$\boldsymbol{{\cal H}}$ with the norm induced by
the scalar product:
\begin{equation}
\label{a4}
{<\hat{\phi}, \hat{\psi}>}_p = \int\limits_{-\infty}^{\infty}
e^{2(p-1)|x|} \sum\limits_{q=0}^p D^q \overline{\hat{\phi}} (x) D^q
\hat{\psi} (x)\;dx \;\;\;;\;\;\;p=1,2,...
\end{equation}
where $dx=dx_1\;dx_2...dx_n$

If we take the conventional scalar product:
\begin{equation}
\label{a5}
<\hat{\phi}, \hat{\psi}> = \int\limits_{-\infty}^{\infty}
\overline{\hat{\phi}}(x) \hat{\psi}(x)\;dx
\end{equation}
then $\boldsymbol{{\cal H}}$, completed with (\ref{a5}), is the Hilbert space
$\boldsymbol{H}$
of square integrable functions.

By definition, the space of continuous linear functionals defined on
$\boldsymbol{{\cal H}}$ is the space
$\boldsymbol{{\Lambda}_{\infty}}$ of the distributions of the exponential
type ( ref.\cite{tp7} ).

The Fourier transform of a distribution of exponential type $\hat{F}$ is given by
(see \cite{tp6, tp7}):
\[F(k)=\int\limits_{-\infty}^{\infty}
H[\Im(k)]H[\Re(x)-H[-\Im(k)]H[-\Re(x)]\hat{F}(x)
e^{ikx}\;dx=\]
\begin{equation}
\label{a6}
H[\Im(k)]\int\limits_0^{\infty}\hat{F}(x)e^{ikx}-
H[-\Im(k)]\int\limits_{-\infty}^0\hat{F}(x)e^{ikx}
\end{equation}
where $F$ is the corresponding tempered ultradistribution
(see the next subsection).

The triplet
\begin{equation}
\label{a7}
\textgoth{\Large{H}}=
\boldsymbol{(}\boldsymbol{{\cal H}},\boldsymbol{H},
\boldsymbol{{\Lambda}_{\infty}} \boldsymbol{)}
\end{equation}
is a Rigged Hilbert Space ( or a Guelfand's triplet \cite{tp12} ).

Moreover, we have: $\boldsymbol{{\cal H}}\subset\boldsymbol{{\cal S}}
\subset\boldsymbol{H}\subset\boldsymbol{{\cal S}^{'}}\subset
\boldsymbol{{\Lambda}_{\infty}}$, where $\boldsymbol{{\cal S}}$ is the
Schwartz space of rapidly decreasing test functions (ref\cite{sch}).

Any Rigged Hilbert Space
$\textgoth{\Large{G}}=\boldsymbol{(}\boldsymbol{\Phi},
\boldsymbol{H},\boldsymbol{{\Phi}^{'}}\boldsymbol{)}$
has the fundamental property that a linear and symmetric operator
on $\boldsymbol{\Phi}$, which admits an extension to a self-adjoint
operator in
$\boldsymbol{H}$, has a complete set of generalized eigenfunctions
in $\boldsymbol{{\Phi}^{'}}$ with real eigenvalues.

\subsection*{Tempered Ultradistributions}

The Fourier transform of a function $\hat{\phi}\in \boldsymbol{{\cal H}}$
is
\begin{equation}
\label{a8}
\phi(z)=\frac {1} {2\pi}
\int\limits_{-\infty}^{\infty}\overline{\hat{\phi}}(x)\;e^{iz\cdot x}\;dx
\end{equation}
Here $\phi(z)$ is entire analytic and rapidly decreasing on straight lines
parallel to the real axis. We call $\boldsymbol{{\EuFrak H}}$
the set of all such functions.
\begin{equation}
\label{a9}
\boldsymbol{{\EuFrak H}}={\cal F}\left\{\boldsymbol{{\cal H}}\right\}
\end{equation}
It is a $\boldsymbol{{\cal Z}\{M_p\}}$
countably normed and complete space ( ref.\cite{gue} ), with:
\begin{equation}
\label{a10}
M_p(z)= (1+|z|)^p
\end{equation}
$\boldsymbol{{\EuFrak H}}$ is a nuclear space defined with the norms:
\begin{equation}
\label{a11}
{\|\phi\|}_{pn} = \sup_{z\in V_n} {\left(1+|z|\right)}^p
|\phi (z)|
\end{equation}
where $V_k=\{z=(z_1,z_2,...,z_n)\in\boldsymbol{{\mathbb{C}}^n}:
\mid Im z_j\mid\leqq k, 1\leqq j \leqq n\}$

We can define the habitual scalar product:
\begin{equation}
\label{a12}
<\phi (z), \psi (z)>=\int\limits_{-\infty}^{\infty}
\phi(z) {\psi}_1(z)\;dz =
\int\limits_{-\infty}^{\infty} \overline{\hat{\phi}}(x)
\hat{\psi}(x)\;dx
\end{equation}
where:
\[{\psi}_1(z)=\int\limits_{-\infty}^{\infty}
\hat{\psi}(x)\; e^{-iz\cdot x}\;dx\]
and $dz=dz_1\;dz_2...dz_n$

By completing $\boldsymbol{{\EuFrak H}}$ with the norm induced by (\ref{a12})
we obtain the Hilbert space of square integrable functions.

The dual of $\boldsymbol{{\EuFrak H}}$ is the space
$\boldsymbol{{\cal U}}$ of tempered ultradistributions
( ref.\cite{tp6, tp7} ). Namely, a tempered ultradistribution is
a continuous linear functional defined on the space
$\boldsymbol{{\EuFrak H}}$ of entire
functions rapidly decreasing on straight lines parallel to the real axis.

The set
$\textgoth{\Large{U}}=
\boldsymbol{({\EuFrak H},H,{\cal U})}$ is also a Rigged Hilbert Space.

Moreover, we have: $\boldsymbol{{\EuFrak H}}\subset\boldsymbol{{\cal S}}
\subset\boldsymbol{H}\subset\boldsymbol{{\cal S}^{'}}\subset
\boldsymbol{{\cal U}}$.

$\boldsymbol{{\cal U}}$ can also be characterized in the following way
( ref.\cite{tp7} ): let $\boldsymbol{{\cal A}_{\omega}}$ be the space of
all functions $F(z)$ such that:

${\boldsymbol{A)}}$
$F(z)$ is analytic on the set $\{z\in \boldsymbol{{\mathbb{C}}^n} :
|Im(z_1)|>p, |Im(z_2)|>p,...,|Im(z_n)|>p\}$.

${\boldsymbol{B)}}$
$F(z)/z^p$ is bounded continuous  in
$\{z\in \boldsymbol{{\mathbb{C}}^n} :|Im(z_1)|\geqq p,|Im(z_2)|\geqq p,
...,|Im(z_n)|\geqq p\}$,
where $p=0,1,2,...$ depends on $F(z)$.

Let $\boldsymbol{\Pi}$ be the set of all $z$-dependent pseudo-polynomials,
$z\in \boldsymbol{{\mathbb{C}}^n}$.
Then $\boldsymbol{{\cal U}}$ is the quotient space:

${\boldsymbol{C)}}$
$\boldsymbol{{\cal U}}=\boldsymbol{{\cal A}_{\omega}/\Pi}$

By a pseudo-polynomial we denote a function of $z$ of the form $\;\;$\\
$\sum_s z_j^s G(z_1,...,z_{j-1},z_{j+1},...,z_n)$ with
$G(z_1,...,z_{j-1},z_{j+1},...,z_n)\in\boldsymbol{{\cal A}_{\omega}}$

Due to these properties it is possible to represent any ultradistribution
as ( ref.\cite{tp7} ):
\begin{equation}
\label{a13}
F(\phi)=<F(z), \phi(z)>=\oint\limits_{\Gamma} F(z) \phi(z)\;dz
\end{equation}
where $\Gamma={\Gamma}_1\cup{\Gamma}_2\cup ...{\Gamma}_n$
and where the path ${\Gamma}_j$ runs parallel to the real axis from
$-\infty$ to $\infty$ for $Im(z_j)>\zeta$, $\zeta>p$ and back from
$\infty$ to $-\infty$ for $Im(z_j)<-\zeta$, $-\zeta<-p$.
( $\Gamma$ surrounds all the singularities of $F(z)$ ).

Formula (\ref{a13}) will be our fundamental representation for a tempered
ultradistribution. Sometimes use will be made of ``Dirac Formula''
for ultradistributions ( ref.\cite{tp6} ):
\begin{equation}
\label{a14}
F(z)=\frac {1} {(2\pi i)^n}\int\limits_{-\infty}^{\infty}
\frac {f(t)} {(t_1-z_1)(t_2-z_2)...(t_n-z_n)}\;dt
\end{equation}
where the ``density'' $f(t)$ is the cut of $F(z)$ along the real
axis and satisfy:\begin{large}                 \end{large}
\begin{equation}
\label{a15}
\oint\limits_{\Gamma} F(z) \phi(z)\;dz =
\int\limits_{-\infty}^{\infty} f(t) \phi(t)\;dt
\end{equation}
While $F(z)$ is analytic on $\Gamma$, the density $f(t)$ is in
general singular, so that the r.h.s. of (\ref{a15}) should be interpreted
in the sense of distribution theory.

Another important property of the analytic representation is the fact
that on $\Gamma$, $F(z)$ is bounded by a power of $z$ ( ref.\cite{tp7} ):
\begin{equation}
\label{a16}
|F(z)|\leq C|z|^p
\end{equation}
where $C$ and $p$ depend on $F$.

The representation (\ref{a15}) implies that the addition of a\\
pseudo-polynomial $P(z)$ to $F(z)$ do not alter the ultradistribution:
\[\oint\limits_{\Gamma}\{F(z)+P(z)\}\phi(z)\;dz=
\oint\limits_{\Gamma} F(z)\phi(z)\;dz+\oint\limits_{\Gamma}
P(z)\phi(z)\;dz\]
But:
\[\oint\limits_{\Gamma} P(z)\phi(z)\;dz=0\]
as $P(z)\phi(z)$ is entire analytic in some of the variables $z_j$
( and rapidly decreasing ),
\begin{equation}
\label{a17}
\therefore \;\;\;\;\oint\limits_{\Gamma} \{F(z)+P(z)\}\phi(z)\;dz=
\oint\limits_{\Gamma} F(z)\phi(z)\;dz
\end{equation}
The inverse Fourier transform of (\ref{a6}) is given by:
\begin{equation}
\label{a18}
\hat{F}(x)=\frac {1} {2\pi}\oint\limits_{\Gamma}F(k)
e^{-ikx}\;dk =\int\limits_{-\infty}^{\infty}f(k)e^{-ikx}\;dx
\end{equation}

\setcounter{section}{2}

\section*{Appendix B}

\setcounter{equation}{0}

\subsection*{Ultradistributions of Exponential Type}

Consider the Schwartz space of rapidly decreasing test functions ${\cal S}$.
Let ${\Lambda}_j$ be the region of the complex plane defined as:
\begin{equation}
\label{b1}
{\Lambda}_j=\left\{z\in\boldsymbol{\mathbb{C}} :
|\Im(z)|< j : j\in\boldsymbol{\mathbb{N}}\right\}
\end{equation}
According to ref.\cite{tp6,tp8} be the space of test functions $\hat{\phi}\in
{\textgoth{\large{V}}}_j$ is
constituted by the ser of all entire analytic functions of ${\cal S}$ for which
\begin{equation}
\label{b2}
||\hat{\phi} ||_j=\max_{k\leq j}\left\{\sup_{z\in{\Lambda}_j}\left[e^{(j|\Re (z)|)}
|{\hat{\phi}}^{(k)}(z)|\right]\right\}
\end{equation}
is finite.\\
The space $\textgoth{\large{Z}}$ is then defined as:
\begin{equation}
\label{b3}
\textgoth{\large{Z}} =\bigcap_{j=0}^{\infty} {\textgoth{\large{V}}}_j
\end{equation}
It is a complete countably normed space with the topology generated by
the set of semi-norms $\{||\cdot ||_j\}_{j\in \mathbb{N}}$.
The topological dual of $\textgoth{\large{Z}}$, denoted by
$\textgoth{\large{B}}$, is by definition the space of ultradistributions of exponential
type (ref.\cite{tp6,tp8}).
Let $\textgoth{S}$ be the space of rapidly decreasing sequences. According to
ref.\cite{tp12} $\textgoth{S}$ is a nuclear space. We consider now the space of
sequences $\textgoth{P}$ generated by the Taylor development of
$\hat{\phi}\in\textgoth{\large{Z}}$
\begin{equation}
\label{b4}
\textgoth{P}=\left\{\textgoth{Q} : \textgoth{Q}
\left(\hat{\phi}(0),{\hat{\phi}}^{'}(0),\frac {{\hat{\phi}}^{''}(0)} {2},...,
\frac {{\hat{\phi}}^{(n)}(0)} {n!},...\right) : \hat{\phi}\in\textgoth{Z}\right\}
\end{equation}
The norms that define the topology of $\textgoth{P}$ are given by:
\begin{equation}
\label{b5}
||\hat{\phi} ||^{'}_p=\sup_n \frac {n^p} {n} |{\hat{\phi}}^n(0)|
\end{equation}
$\textgoth{P}$ is a subspace of $\textgoth{S}$ and as consequence is a nuclear space.
The norms $||\cdot ||_j$ and $||\cdot ||^{'}_p$ are equivalent, the correspondence
\begin{equation}
\label{b6}
\textgoth{\large{Z}}\Longleftrightarrow \textgoth{P}
\end{equation}
is an isomorphism and therefore $\textgoth{Z}$ is a countably normed nuclear space.
We define now the set of scalar products
\[<\hat{\phi}(z),\hat{\psi}(z)>_n=\sum\limits_{q=0}^n\int\limits_{-\infty}^{\infty}e^{2n|z|}
\overline{{\hat{\phi}}^{(q)}}(z){\hat{\psi}}^{(q)}(z)\;dz=\]
\begin{equation}
\label{b7}
\sum\limits_{q=0}^n\int\limits_{-\infty}^{\infty}e^{2n|x|}
\overline{{\hat{\phi}}^{(q)}}(x){\hat{\psi}}^{(q)}(x)\;dx
\end{equation}
This scalar product induces the norm
\begin{equation}
\label{b8}
||\hat{\phi}||_n^{''}=[<\hat{\phi}(x),\hat{\phi}(x)>_n]^{\frac {1} {2}}
\end{equation}
The norms $||\cdot ||_j$ and $||\cdot ||^{''}_n$ are equivalent, and therefore
$\textgoth{\large{Z}}$ is a countably hilbertian nuclear space.
Thus, if we call now ${\textgoth{\large{Z}}}_p$ the completion of
$\textgoth{\large{Z}}$ by the norm $p$ given in (\ref{b8}), we have:
\begin{equation}
\label{b9}
\textgoth{\large{Z}}=\bigcap_{p=0}^{\infty}{\textgoth{\large{Z}}}_p
\end{equation}
where
\begin{equation}
\label{b10}
{\textgoth{\large{Z}}}_0=\boldsymbol{H}
\end{equation}
is the Hilbert space of square integrable functions.\\
As a consequence the triplet
\begin{equation}
\label{b11}
\textgoth{\Large{U}}=\boldsymbol{(}\textgoth{\large{Z}},
\boldsymbol{H}, \textgoth{\large{B}}\boldsymbol{)}
\end{equation}
is also a Guelfand's triplet.

$\textgoth{\large{B}}$ can also be characterized in the following way
( refs.\cite{tp6},\cite{tp8} ): let ${\textgoth{E}}_{\omega}$ be the space of
all functions $\hat{F}(z)$ such that:

${\boldsymbol{A)}}$
$\hat{F}(z)$ is an analytic function for $\{z\in \boldsymbol{\mathbb{C}} :
|Im(z)|>p\}$.

${\boldsymbol{B)}}$-
$\hat{F}(z)e^{-p|\Re(z)|}/z^p$ is a bounded continuous function  in
$\{z\in \boldsymbol{\mathbb{C}} :|Im(z)|\geqq p\}$,
where $p=0,1,2,...$ depends on $\hat{F}(z)$.

Let $\textgoth{N}$ be:
$\textgoth{N}=\{\hat{F}(z)\in{\textgoth{E}}_{\omega} :\hat{F}(z)\; \rm{is\; entire\; analytic}\}$.
Then $\textgoth{\large{B}}$ is the quotient space:

${\boldsymbol{C)}}$-
$\textgoth{\large{B}}={\textgoth{E}}_{\omega}/\textgoth{N}$

Due to these properties it is possible to represent any ultradistribution
of exponential type as ( ref.\cite{tp6,tp8} ):
\begin{equation}
\label{b12}
\hat{F}(\hat{\phi})=<\hat{F}(z), \hat{\phi}(z)>=\oint\limits_{\Gamma} \hat{F}(z) \hat{\phi}(z)\;dz
\end{equation}
where the path ${\Gamma}$ runs parallel to the real axis from
$-\infty$ to $\infty$ for $Im(z)>\zeta$, $\zeta>p$ and back from
$\infty$ to $-\infty$ for $Im(z)<-\zeta$, $-\zeta<-p$.
( $\Gamma$ surrounds all the singularities of $\hat{F}(z)$ ).

Formula (\ref{b12}) will be our fundamental representation for a
ultradistribution of exponential type.
The ``Dirac Formula''
for ultradistributions of exponential type is( ref.\cite{tp6, tp8} ):
\begin{equation}
\label{b13}
\hat{F}(z)\equiv\frac {1} {2\pi i}\int\limits_{-\infty}^{\infty}
\frac {\hat{f}(t)} {t-z}\;dt\equiv
\frac {\cosh(\lambda z)} {2\pi i}\int\limits_{-\infty}^{\infty}
\frac {\hat{f}(t)} {(t-z)\cosh(\lambda t)}\;dt
\end{equation}
where the ``density'' $\hat{f}(t)$ is such that
\begin{equation}
\label{b14}
\oint\limits_{\Gamma} \hat{F}(z) \hat{\phi}(z)\;dz =
\int\limits_{-\infty}^{\infty} \hat{f}(t) \hat{\phi}(t)\;dt
\end{equation}
(\ref{b13}) should be used carefully.
While $\hat{F}(z)$ is analytic function on $\Gamma$, the density $\hat{f}(t)$ is in
general singular, so that the right hand side of (\ref{b14}) should be interpreted again
in the sense of distribution theory.

Another important property of the analytic representation is the fact
that on $\Gamma$, $\hat{F}(z)$ is bounded by a exponential and a power of $z$
( ref.\cite{tp6,tp8} ):
\begin{equation}
\label{b15}
|\hat{F}(z)|\leq C|z|^pe^{p|\Re(z)|}
\end{equation}
where $C$ and $p$ depend on $\hat{F}$.

The representation (\ref{b12}) implies that the addition of any entire function
$\hat{G}(z)\in\textgoth{N}$ to $\hat{F}(z)$ does not alter the ultradistribution:
\[\oint\limits_{\Gamma}\{\hat{F}(z)+\hat{G}(z)\}\hat{\phi}(z)\;dz=
\oint\limits_{\Gamma} \hat{F}(z)\hat{\phi}(z)\;dz+\oint\limits_{\Gamma}
\hat{G}(z)\hat{\phi}(z)\;dz\]
But:
\[\oint\limits_{\Gamma} \hat{G}(z)\hat{\phi}(z)\;dz=0\]
as $\hat{G}(z)\hat{\phi}(z)$ is an entire analytic function,
\begin{equation}
\label{b16}
\therefore \;\;\;\;\oint\limits_{\Gamma} \{\hat{F}(z)+\hat{G}(z)\}\hat{\phi}(z)\;dz=
\oint\limits_{\Gamma} \hat{F}(z)\hat{\phi}(z)\;dz
\end{equation}

Another very important property of $\textgoth{\large{B}}$ is that
$\textgoth{\large{B}}$ is reflexive under the Fourier transform:
\begin{equation}
\label{b17}
\textgoth{\large{B}}={\cal F}_c\left\{\textgoth{\large{B}}\right\}=
{\cal F}\left\{\textgoth{\large{B}}\right\}
\end{equation}
where the complex Fourier transform $F(k)$ of $\hat{F}(z)\in\textgoth{\large{B}}$
is given by:
\[F(k)=H[\Im(k)]\int\limits_{{\Gamma}_+}\hat{F}(z)e^{ikz}\;dz-
H[-\Im(k)]\int\limits_{{\Gamma}_{-}}\hat{F}(z)e^{ikz}\;dz=\]
\[\oint\limits_{\Gamma}\{H[\Im(k)H[\Re(z)]-H[-\Im(k)H[-\Re(z)]\}
\hat{F}(z)e^{ikz}\;dz=\]
\begin{equation}
\label{b18}
H[\Im(k)]\int\limits_0^{\infty}\hat{f}(x)e^{ikx}\;dx-
H[-\Im(k)]\int\limits_{-\infty}^0\hat{f}(x)e^{ikx}\;dx
\end{equation}
Here ${\Gamma}_+$ is the part of $\Gamma$ with $\Re(z)\geq 0$ and
${\Gamma}_{-}$ is the part of $\Gamma$ with $\Re(z)\leq 0$
Using (\ref{b18}) we can interpret Dirac's Formula as:
\begin{equation}
\label{b19}
F(k)\equiv\frac {1} {2\pi i}\int\limits_{-\infty}^{\infty}
\frac {f(s)} {s-k}\; ds\equiv{\cal F}_c\left\{{\cal F}^{-1}\left\{f(s)\right\}\right\}
\end{equation}
The inverse Fourier transform corresponding to (\ref{b19}) is
given by:
\begin{equation}
\label{b20}
\hat{F}(z)=\frac {1} {2\pi}
\oint\limits_{\Gamma}\{H[\Im(z)]H[-\Re(k)]-H[-\Im(z)]H[\Re(k)]\}
F(k)e^{-ikz}\;dk
\end{equation}

The treatment for ultradistributions of exponential type defined on
${\boldsymbol{\mathbb{C}}}^n$ is similar to the case of one variable.
Thus let ${\Lambda}_j$ be given as
\begin{equation}
\label{b21}
{\Lambda}_j=\left\{z=(z_1, z_2,...,z_n)\in{\boldsymbol{\mathbb{C}}}^n :
|\Im(z_k)|\leq j\;\;\;1\leq k\leq n\right\}
\end{equation}
and
\begin{equation}
\label{b22}
||\hat{\phi} ||_j=\max_{k\leq j}\left\{\sup_{z\in{\Lambda}_j}\left[
e^{j\left[\sum\limits_{p=1}^n|\Re(z_p)|\right]}\left| D^{(k)}\hat{\phi}(z)\right|\right]\right\}
\end{equation}
where $D^{(k)}={\partial}^{(k_1)}{\partial}^{(k_2)}\cdot\cdot\cdot{\partial}^{(k_n)}\;\;\;\;
k=k_1+k_2+\cdot\cdot\cdot+k_n$

${\textgoth{\large{B}}}^n$ is characterized as follows. Let
${\textgoth{E}}^n_{\omega}$ be the space of all functions $\hat{F}(z)$ such that:

${\boldsymbol{A^{'})}}$
$\hat{F}(z)$ is analytic for\\ $\{z\in \boldsymbol{{\mathbb{C}}^n} :
|Im(z_1)|>p, |Im(z_2)|>p,...,|Im(z_n)|>p\}$.

${\boldsymbol{B^{'})}}$
$\hat{F}(z)e^{-\left[p\sum\limits_{j=1}^n|\Re(z_j)|\right]}/z^p$
is bounded continuous  in
$\{z\in \boldsymbol{{\mathbb{C}}^n} :|Im(z_1)|\geqq p,|Im(z_2)|\geqq p,
...,|Im(z_n)|\geqq p\}$,
where $p=0,1,2,...$ depends on $\hat{F}(z)$.

Let ${\textgoth{N}}^n$ be:
${\textgoth{N}}^n=\left\{\hat{F}(z)\in{\textgoth{E}}^n_{\omega} :\hat{F}(z)\;\right.$
is entire analytic function at minus in one of the variables $\left. z_j\;\;\;1\leq j\leq n\right\}$
Then ${\textgoth{\large{B}}}^n$ is the quotient space:

${\boldsymbol{C^{'})}}$
${\textgoth{\large{B}}}^n={\textgoth{E}}^n_{\omega}/{\textgoth{N}}^n$
We have now
\begin{equation}
\label{b23}
\hat{F}(\hat{\phi})=<\hat{F}(z), \hat{\phi}(z)>=\oint\limits_{\Gamma} \hat{F}(z) \hat{\phi}(z)\;
dz
\end{equation}
where $\Gamma={\Gamma}_1\cup{\Gamma}_2\cup ...{\Gamma}_n$
and where the path ${\Gamma}_j$ runs parallel to the real axis from
$-\infty$ to $\infty$ for $Im(z_j)>\zeta$, $\zeta>p$ and back from
$\infty$ to $-\infty$ for $Im(z_j)<-\zeta$, $-\zeta<-p$.
(Again the path $\Gamma$ surrounds all the singularities of $\hat{F}(z)$ ).
The n-dimensional Dirac's Formula is now
\begin{equation}
\label{b24}
\hat{F}(z)=\frac {1} {(2\pi i)^n}\int\limits_{-\infty}^{\infty}
\frac {\hat{f}(t)} {(t_1-z_1)(t_2-z_2)...(t_n-z_n)}\;dt
\end{equation}
and the ``density'' $\hat{f}(t)$ is such that
\begin{equation}
\label{b25}
\oint\limits_{\Gamma} \hat{F}(z)\hat{\phi}(z)\;dz =
\int\limits_{-\infty}^{\infty} \hat{f}(t) \hat{\phi}(t)\;dt
\end{equation}
The modulus of $\hat{F}(z)$ is bounded by
\begin{equation}
\label{b26}
|\hat{F}(z)|\leq C|z|^p e^{\left[p\sum\limits_{j=1}^n|\Re(z_j)|\right]}
\end{equation}
where $C$ and $p$ depend on $\hat{F}$.

\setcounter{section}{3}

\section*{Appendix C}

\setcounter{equation}{0}

\subsection*{Fractional derivative}

According to \cite{tp1} the fractional derivative of a
distribution of exponential type $\hat{F}(x)$ is given by
\begin{equation}
\label{c1}
\frac {d^{\lambda}\hat{F}(x)} {dx^{\lambda}}=\frac {1} {2\pi}\oint\limits_{\Gamma}
(-ik)^{\lambda}
F(k)
e^{-ik x}\;dk+
\oint\limits_{\Gamma}(-ik)^{\lambda}a(k)
e^{-ik x}\;dk
\end{equation}
Where $a(k)$ is entire analytic and rapidly decreasing.
If $\lambda=-1$, $d^{\lambda}/dx^{\lambda}$ is the inverse of the derivative
(an integration). In this case the second term of the right side of (\ref{c1})
gives a primitive of $\hat{f}(x)$. Using Cauchy's theorem the additional
term is
\begin{equation}
\label{c2}
\oint\limits_{\Gamma} \frac {a(k)} {k}e^{-ik x}\;dk=
2\pi a(0)
\end{equation}
Of course, an integration should give a primitive plus an arbitrary constant.
Analogously when $\lambda=-2$ (a double iterated integration) we have
\begin{equation}
\label{c3}
\oint\limits_{\Gamma} \frac {a(k)} {{k}^2}e^{-ik x}\;dk=
\gamma+\delta x
\end{equation}
where $\gamma$ and $\delta$ are arbitrary constants.

For a ultradistribution of exponential type we have for the fractional
derivative:
\[\frac {{\partial}^{\lambda}\hat{F}(z)} {\partial z^{\lambda}}=\frac {1} {2\pi}
\oint\limits_{\Gamma}\{H[\Im(z)H[-\Re(k)]-H[-\Im(z)H[\Re(k)]\}
(-ik)^{\lambda}F(k)e^{-ikz}\;dk+\]
\begin{equation}
\label{c4}
\oint\limits_{\Gamma}\{H[\Im(z)H[-\Re(k)]-H[-\Im(z)H[\Re(k)]\}
(-ik)^{\lambda}a(k)e^{-ikz}\;dk
\end{equation}
where $a(k)\in\textgoth{\large{Z}}$.
This fractional derivative behaves similarly to the above-defined
for distributions of exponential type.

Unlike all other definitions of fractional derivative,
(\ref{c1}) and (\ref{c4}) are defined for all values
of $\lambda$, real o complex.
Furthermore, are the only known definitions that
unify derivation and integration in a single operation.

\setcounter{section}{4}

\section*{Appendix D}

\setcounter{equation}{0}

\subsection*{Some useful formulas related to the Hypergeometric Function}

According to the result given in \cite{gra2} we can obtain:
\[\int\limits_0^t\frac {t^{'n}} {(x+at^{'}\pm i0)^{\lambda n +1}}\;dt^{'}=
\frac {t^{n+1}} {(x\pm i0)^{\lambda n+1}}B(1,n+1)\times\]
\begin{equation}
\label{d1}
F\left(\lambda n +1, n+1,n+2;-\frac {at} {x\pm
i0}\right).
\end{equation}
Using the transformation formula given in \cite{gra3}
for the hypergeometric function
\[F(\lambda+1,2;3;z)=\frac {2\Gamma(1-\lambda)} {\Gamma(2-\lambda)}
(-1)^{\lambda+1}z^{-\lambda-1}
F\left(\lambda+1,\lambda-1;\lambda;\frac {1} {z}\right)+\]
\begin{equation}
\label{d2}
\frac {2\Gamma(\lambda-1)} {\Gamma(\lambda+1)} z^{-2}
F\left(2,0;2-\lambda;\frac {1} {z}\right),
\end{equation}
with the particular value
\begin{equation}
\label{d3}
F(a,0;c;z)=1.
\end{equation}
we obtain the expression:
\[F(\lambda+1,2;3;z)=\frac {2\Gamma(1-\lambda)} {\Gamma(2-\lambda)}
(-1)^{\lambda+1}z^{-\lambda-1}
F\left(\lambda+1,\lambda-1;\lambda;\frac {1} {z}\right)+\]
\begin{equation}
\label{d4}
\frac {2\Gamma(\lambda-1)} {\Gamma(\lambda+1)}z^{-2}
\end{equation}
Now by recourse to  the transformation formula \cite{gra4} we have:
\begin{equation}
\label{d5}
F\left(\lambda+1,\lambda-1;\lambda;\frac {1} {z}\right)=
\left(1-\frac {1} {z}\right)^{-\lambda}F\left(-1,1;\lambda;\frac {1}
{z}\right),
\end{equation}
or equivalently:
\begin{equation}
\label{d6}
F\left(\lambda+1,\lambda-1;\lambda;\frac {1} {z}\right)=
\frac {z^{\lambda}} {(z-1)^{\lambda}}\left(\frac {\lambda z-1}
{\lambda z}\right).
\end{equation}
Thus, we get, finally,
\begin{equation}
\label{d7}
F(\lambda+1,2;3;z)=\frac {2} {\lambda(\lambda-1)z^2} \left[1+\frac
{\lambda z-1} {(1-z)^{\lambda}}\right].
\end{equation}

\end{document}